# Flexible graphene/boron nitride nanosheets paper for thermal management of high power electronics

Xiaojuan Tian

**Abstract**

Graphene nanosheets (GNS) paper is widely regarded as a promising candidate for heat dissipation due to its outstanding thermal conductivity. However, the accompanied high electrical conductivity makes it unfavorable for thermal management of high power electronics since it runs a high risk of short circuits. To eliminate the risk from the high electrical conductivity and simultaneously maintain the excellent thermal performance, we introduce boron nitride nanosheets (BNNS) that possess high thermal conductivity but electrical insulation into the GNS paper. The hybrid paper has a much lower electrical conductivity but similar thermal performance compared to the pristine GNS paper. Besides, the flexible hybrid paper exhibits better thermal stability than pure GNS paper. Our results show that the ability of BNNS to change the electrical conductivity of paper without affecting its thermal conductivity is potential for the application of heat management materials with tailored electrical properties.

**Key words**: graphene, boron nitride nanosheets, thermal conductivity, electrical conductivity

## 1. Introduction

The growth of packaging density in advanced electronic circuitry has resulted in a pronounced increasing of the heat generation. Graphene based materials are widely used for thermal management because of the excellent thermal conductivity[1-9]. For example, they are popular thermally conductive fillers in polymer based thermal interface materials[9-16], whose performance is limited by the low thermal conductivity of the polymer[14]. Recently, the thermal management applications of graphene sheets go beyond the polymer composites extending to all carbon papers by applying the acid treated GNS of outstanding electrical, thermal and mechanical properties[17]. Although the produced graphene paper is highly thermally conductive, the accompanied high electrical conductivity limits its applications in thermal management of high power electronics since it may result in short circuits.

Boron nitride nanosheets (BNNS), structurally analogous to graphene, has a high thermal conductivity of 2000 W/mK[18]. At the same time, it is an electrical insulator with a dielectric constant of 3-4[19]. Boron nitride nanosheets is usually produced through liquid phase exfoliation[20, 21], and have been applied in polymer composites with thermal conductivity of ~ 5 W/mK[22, 23].

Here, we combined the acid treated GNS and BNNS to prepare flexible composite. The BNNS break the percolating electrons transfer, while maintaining the high thermal conductivity, thus perfectly suited for thermal management of high power electronics.

## 2. Experimental

BNNS were obtained through liquid phase exfoliation of commercial bulk h-BN powder (Dandong Chemical Engineering Institute, China). In a typical experiment, h-BN powder (500 mg) was sonicated in DI water (200 mL) for 8h. The supernatant was collected after staying overnight.

The GNS were produced by shear-assisted supercritical $CO_2$ exfoliation according to the previous literature[24] and further treated in a combination of alcohol and acid treatments. In a typical experiment, GNS (250 mg) was sonicated in a mixture of alcohol and water (13:7 vol.) for 24h. Then the products were collected through filtration and dried in vacuum oven followed by sonicating in a mixture of nitric acid and sulfuric acid (40 ml, v($HNO_3$):v($H_2SO_4$) is 3:1). The acid solution was diluted into 1L DI water to produce aqueous suspension.

The flexible papers were prepared by filtering the aqueous solution of BNNS and GNS, peeled off and dried in vacuum oven at 75ºC for 8 hours.

Transmission electron microscope (TEM) of BNNS and GNS were conducted on FEI F20. The thermal diffusivity ($\alpha$) of the papers was measured by a LFA Nanoflash 447 system (Netzsch). And the specific heat capacity (Cp) was measured by DSC (200 F3, Netzsch). The thermal conductivity (κ) is calculated according to the equation: $\kappa = \alpha \times \rho \times C_p$, where the ρ is the apparent density. Electrical conductivity was measured by four point method with Kunde KDY-1 system (Guangzhou, P. R. China). Thermal gravimetric analysis (TGA) was conducted in nitrogen at a heating rate of 10 °C min$^{-1}$.

## 3. Results and Discussion

Liquid phase exfoliation is widely used for exfoliating layered materials, such as graphite, BN, and WS$_2$[21]. The DI water exfoliation is effective in exfoliating the layered h-BN without the use of surfactants or organic functionalization[20]. The pristine BN powder and BNNS were investigated under TEM. Fig 1 a, b present the h-BN powder with tens to hundreds layers. After exfoliation in DI water, it is clear from Fig 1 c, d that the exfoliated BNNS are thinner and more transparent with around ten layers, obviously more exfoliated than the pristine h-BN particles. The GNS was prepared by supercritical CO$_2$ exfoliation followed by acid treatment. The product was investigated through SEM and TEM. It is seen that the size if GNS is several micrometers (Fig. 2a). Under TEM and HR-TEM, it is clearly that the GNS applied is much thinner than graphite, which is favorable for thermal performance. There is a small D band in the Raman spectra of the GNS (Fig. 2d), which could attribute to the defects associated with new edges introduced via exfoliation, agreeing well with the reports of liquid phase exfoliated graphene[25]. The I$_{2D}$/I$_G$ is 0.56, indicating the products are multi-layer graphene nanosheets.

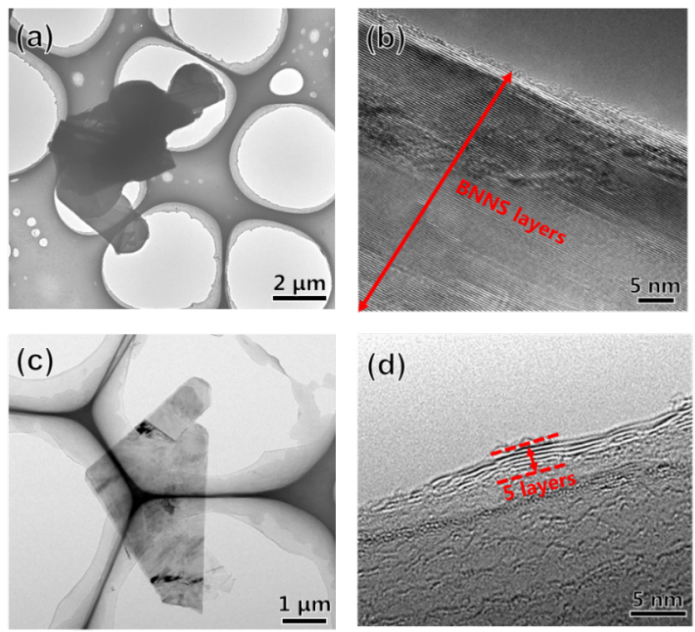

Figure 1

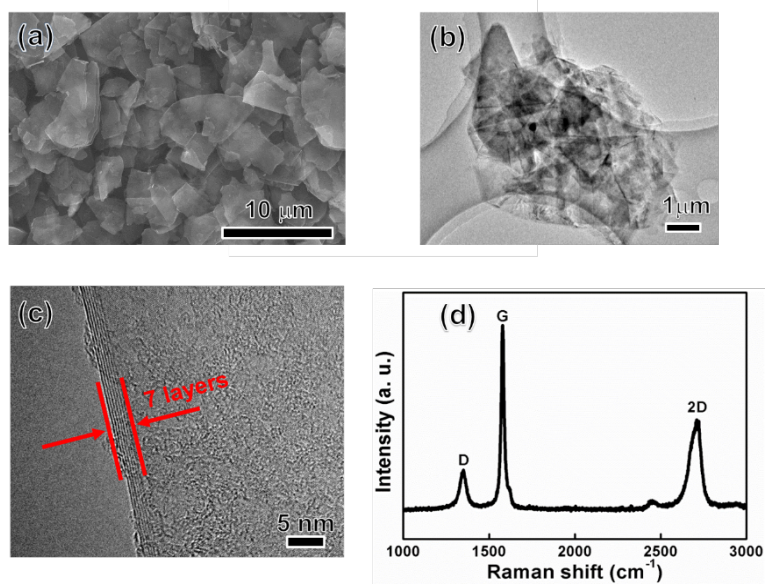

Figure 2

The flexible papers composed of GNS/BNNS were prepared through vacuum filtration. It is worthing noting that the acid treatment of GNS plays an important role for formation of a free-standing film. Thus, the acid treated GNS not only play the role of highly thermally conductive fillers, but also serve as the structural component to reinforce the mechanical strength and flexibility. BNNS are expected to contribute to the pathways of acoustic phonon transmission, while blocking the percolating electrons transfer, as shown in Fig. 3a. The produced paper is uniform and flexible to bend into big angles.(Fig. 3b) The electrical and thermal conductivities of the papers as a function of BNNS weight loadings are presented in Fig. 3c. The pristine acid treated GNS has an electrical conductivity of 81 S/cm. With the introduction of BNNS, the electrical conductivity degrades rapidly. At 75% wt. BNNS, the electrical conductivity is only 3 S/cm, which is similar to that of the 14% wt. GNS/polymer thermal interface materials (2.5~4 S/cm) [14]. At the same time, the thermal conductivity of the BNNS/GNS papers remains ~ 40 W/mK, which is much higher than that of GNS/polymer composites (~5 W/mK) [14]. In terms of the BNNS/GNS paper, both components contribute to the phonon transmission, while the thermal conductivity of the polymer is as low as 0.15 W/mK in the GNS/polymer system[14]. The dramatic drops of electrical conductivity via introduction of BNNS could attribute to the huge gap between the electrical properties of GNS and BNNS. The electrical conductivity of graphene is reported to be 2000 S/cm [26], while the BNNS is insulating. The thermal conductivity of single layer graphene is 5000 W/mK [9]. However, the thermal properties of graphene degrade dramatically with the increase of the number of layers because of the phonon boundary scattering[27]. The thermal conductivity of multi layer graphene is 1000~2000 W/mK[27], which is similar to that of BNNS[18], so that the thermal properties is stable with varying the proportions of BNNS and GNS. To evaluate the utilization of the composites as thermal management materials for high power electronics, we take the ratio of thermal conductivity to electrical conductivity (Fig. 3d). With the increase of BNNS amount, the ratio increases. For the pristine GNS paper, the thermal conductivity divided by electrical conductivity is 0.5, in agreement with the previously reported graphene based papers[17] [28]. When the BNNS weight loading reaches 60%, the ratio of thermal conductivity to electrical conductivity improves to 4, thus more suitable for applications as heat dissipation materials for high power electronics in case the short-circuit risk.

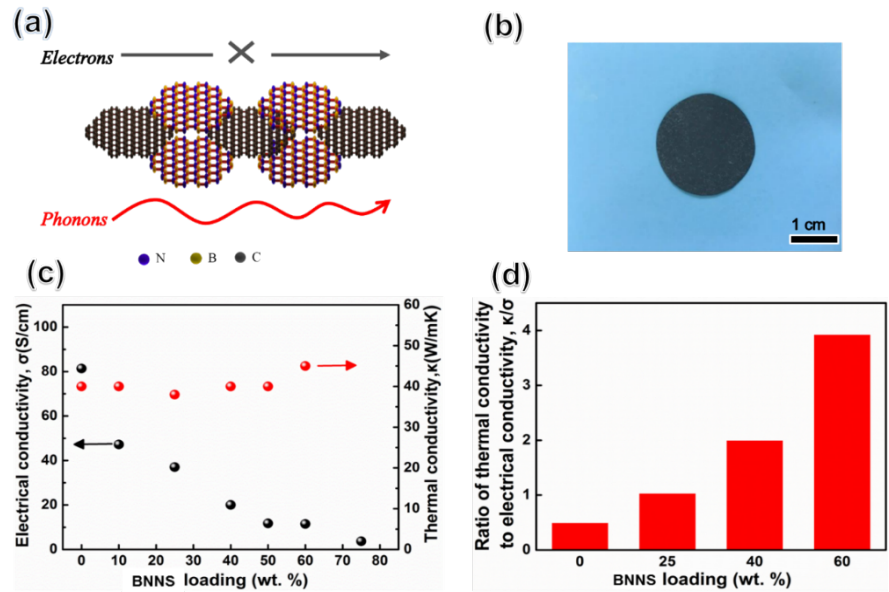

Figure 3

In addition, the thermal stability of the sample is examined by the TGA under $N_2$, as shown in Fig. 4. The pristine GNS paper has the most weight loss with the increase of temperature, which is from the functional groups introduced by the acid treatments. Obviously, the thermal stability becomes much better by increasing the BNNS loading up to 50% wt., which can be explained by the high thermally stability of boron nitride at 1000ºC. Therefore, the introduction of BNNS not only decrease the risk of short-circuit, but also increase the thermal stability of film, making it quite promising for thermal management applications.

## 4. Conclusions

In summary, the novel flexible paper was prepared with BNNS and GNS. It is found that the addition of BNNS can greatly decrease the electrical conductivity, while effectively maintaining the high thermal conductivity, which completely eliminates the risk of short circuits. As the BNNS weight loading increase to 60%, the ratio of thermal conductivity to electrical conductivity is up to 4 which is more suitable for heat dissipation of high power electronics. Moreover, the film has a better thermal stability from the BNNS introduction. Thus the GNS/BNNS flexible paper is promising for thermal management materials with tailored properties of high power electronics.

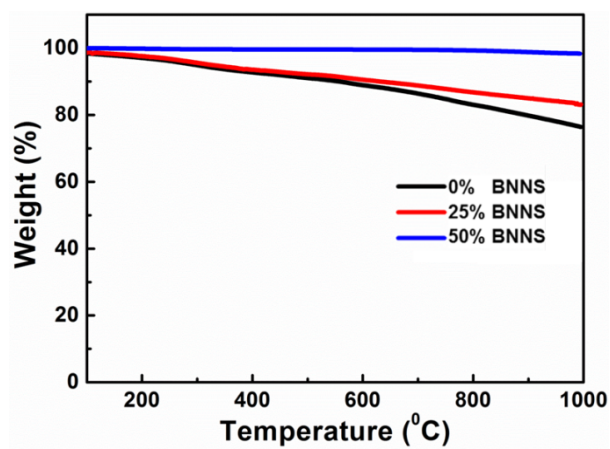

Figure 4

**Acknowledgements**

This work was supported by the National Natural Science Foundation of China (Nos. 21808240 and 21776308).

**Figure captions**

**Fig. 1.** Low magnification (a) and high magnification (b) TEM of h-BN powder. Low magnification (c) and high magnification (d) TEM of exfoliated BNNS.

**Fig. 2.** SEM (a), low magnification (b) and high magnification (c) TEM of GNS. (d) Raman spectra of GNS.

**Fig. 3.** (a) Scheme and (b) photograph of the GNS/BNNS paper. (c) Electrical and thermal conductivities of the samples as a function of BNNS loading. (d) Ratio of thermal to electrical conductivity vs. the BNNS loading.

**Fig. 4.** TGA of GNS/BNNS papers under $N_2$.